\documentclass[sigconf]{acmart}
\AtBeginDocument{%
  }

\copyrightyear{2026}
\acmYear{2026}
\setcopyright{cc}
\setcctype{by}
\acmConference[CNSSE 2026]{2026 6th International Conference on Computer Network Security and Software Engineering}{February 06--08, 2026}{Qingdao, China}
\acmBooktitle{2026 6th International Conference on Computer Network Security and Software Engineering (CNSSE 2026), February 06--08, 2026, Qingdao, China}
\acmDOI{10.1145/3803633.3803665}
\acmISBN{979-8-4007-2195-3/2026/02}

\usepackage{soul}
\usepackage{bm}
\usepackage{xcolor}
\usepackage{booktabs}
\usepackage{microtype}
\usepackage{amsfonts}
\usepackage{xspace}
\usepackage{wasysym}
\usepackage{enumitem,etoolbox}
\usepackage{array}
\usepackage{microtype}
\usepackage{setspace}
\usepackage{multirow}
\usepackage{framed}
\usepackage{caption}
\usepackage{appendix}
\usepackage{flushend}
\usepackage{balance}
\usepackage{graphicx}
\usepackage{amsthm}
\usepackage{amsmath}
\usepackage{amssymb}

\usepackage{bbding}
\usepackage{array}
\usepackage{multirow}
\usepackage{listings}
\usepackage{listings-ext}
\usepackage{color}
\usepackage{tikz}
\usepackage{enumitem}

\setlist[itemize]{leftmargin=*}
\setlist[enumerate]{leftmargin=*}
\newlist{steps}{enumerate}{1}
\setlist[steps, 1]{label = \textbf{RQ\arabic*.}}

\usepackage{hyperref}
\usepackage{url}
\usepackage{xcolor,colortbl,color}
\hypersetup{
 colorlinks=true,
 linkcolor=purple,
 citecolor=violet,
 filecolor=magenta,      
 urlcolor=cyan,
}

\setlist[itemize]{leftmargin=*}
\setlist[enumerate]{leftmargin=*}

\definecolor{rowcolor}{HTML}{ECEFF4}

\setlist[itemize]{leftmargin=*}

\definecolor{dkgreen}{rgb}{0,0.6,0}
\definecolor{gray}{rgb}{0.5,0.5,0.5}
\definecolor{mauve}{rgb}{0.58,0,0.82}

\begin{document}

\title{From Heuristics to Transformers: A Comprehensive Survey of Type Inference from Stripped Binaries}


\author{Hua Zheng}
\orcid{0000-0001-7799-1912}
\authornote{Both authors contributed equally to this research.}
\affiliation{%
  \institution{Guangzhou University of Software}
  \city{Guangzhou}
  \country{China}
}
\email{zhengh@mail.gzus.edu.cn}

\author{Yuhang Guo}
\authornotemark[1]
\orcid{0009-0003-7326-8665}
\affiliation{%
  \institution{Guangzhou Huali College} 
  \city{Guangzhou}
  \country{China}
}
\email{1203515292@qq.com}

\author{Kuanishbay Sadatdiynov}
\orcid{0000-0003-1759-5103}
\authornote{Corresponding author: Cheng Wen and Kuanishbay Sadatdiynov}
\affiliation{%
  \institution{Nukus State Technical University}
  \city{Nukus}
  \country{Uzbekistan}
}
\email{k.sadatdiynov@nukusstu.uz}

\author{Cheng Wen}
\orcid{0000-0003-1826-6213}
\authornotemark[2]
\affiliation{%
  \institution{Xidian University}
  \city{Xi'an}
  \country{China}
}
\email{wencheng@xidian.edu.cn}

\author{Muhammad Sadiq}
\orcid{0000-0003-2199-3702}
\affiliation{%
  \institution{Shenzhen University of Information Technology}
  \city{Shenzhen}
  \country{China}
}
\email{sadiq@sziit.edu.cn}

\author{Dugang Liu}
\orcid{0000-0003-3612-709X}
\affiliation{%
  \institution{Shenzhen University}
  \city{Shenzhen}
  \country{China}
}
\email{liudugang@szu.edu.cn}

\author{Jawwad Ahmed Shamsi}
\orcid{0000-0001-6813-2673}
\affiliation{%
  \institution{National University of Computer and Emerging Sciences Karachi}
  \city{Karachi}
  \country{Pakistan}
}
\email{Jawwad.shamsi@nu.edu.pk}

\author{Anam Qureshi}
\orcid{0000-0003-0491-5788}
\affiliation{%
  \institution{National University of Computer and Emerging Sciences}
  \city{Karachi}
  \country{Pakistan}
}
\email{anam.qureshi@nu.edu.pk}

\renewcommand{\shortauthors}{Hua et al.}

\begin{abstract}
  The recovery of high-level type information from stripped binaries—executables devoid of symbol tables and debugging information—is a cornerstone of software reverse engineering, vulnerability analysis, and decompilation.
This survey tracks the evolution of binary type inference from early rule-based heuristics and static analysis to modern deep learning architectures.
We analyze the shift from "duck typing" and constraint-solving techniques (\textit{e.g.}, BITY, BinSub) to context-aware neural models (\textit{e.g.}, EKLAVYA, CATI) and finally to state-of-the-art Transformer and Graph Neural Network (GNN) architectures (\textit{e.g.}, SeeType, TYGR). 
We identify core challenges, including optimization-induced semantics loss and structural type recovery, and propose future research directions in neuro-symbolic inference.
\end{abstract}




\maketitle

\section{Introduction}
The rapid proliferation of software in modern society has made the ability to analyze and understand executable binaries a critical necessity for cybersecurity~\cite{javed2025binary,huang2024binary,wang2020typestate}. Whether for the purpose of vulnerability discovery~\cite{xin2025irhunter,wen2020memlock,wen2022controlled} and malware analysis~\cite{xu2017effective} in legacy systems, or verifying the security of closed-source third-party components, the field of binary reverse engineering serves as the primary line of defense. However, the task of understanding a program without access to its source code is an inherently asymmetric struggle~\cite{shao2022survey}.

Fundamentally, compilation is a lossy transformation. High-level programming languages provide developers with powerful abstractions—most notably data types—that impart semantic meaning to raw data~\cite{soni2025benchmarking}. During compilation, these abstractions are stripped away. What was once a ``\texttt{struct student\_record}'' or a ``\texttt{char* buffer}'' is reduced to a generic sequence of loads and stores into fixed-size registers and memory offsets. In "stripped" binaries, the loss is even more severe: symbol tables and debugging information (such as DWARF) are removed to minimize file size and protect proprietary logic, leaving behind a ``semantic desert'' of raw assembly instructions.

\textbf{The Role of Type Inference.}
Binary Type Inference (BTI) is the process of reconstructing these high-level types from machine code~\cite{xu2017learning,xu2021extracting}. It is perhaps the most critical component of a modern decompiler. Precise type information acts as the "glue" that allows a decompiler to transform flat assembly into structured, human-readable C-like code. Without it, pointers cannot be distinguished from integers, and the boundaries of complex data structures (\textit{e.g.}, arrays, structures, and unions) remain invisible~\cite{zeng2018debugging}. Moreover, type inference is a prerequisite for advanced security applications, such as Control-Flow Integrity (CFI)~\cite{abadi2009control}, where type-signature matching is used to restrict the possible targets of indirect function calls, thereby mitigating return-oriented programming (ROP) attacks.

\textbf{From Heuristics to Deep Learning.}
For decades, type inference was the domain of manual expert knowledge. Classic tools like IDA Pro~\cite{ferguson2008reverse} and Ghidra~\cite{eagle2020ghidra} relied on rule-based heuristics, often referred to as "Duck Typing"—the logic that if a variable is used as an index into memory, it must be an array or a pointer. While effective for simple cases, these heuristics are brittle. They struggle with aggressive compiler optimizations (\textit{e.g.}, \texttt{-O3}) that reorder code and reuse registers for multiple variables of different types. Furthermore, the sheer variety of instruction set architectures (ISAs)—from x86 and ARM to MIPS and AArch64—makes it impossible for human analysts to keep pace by writing manual rules for every new compiler idiom.

To address these limitations, the research community has recently pivoted toward data-driven approaches. The evolution of BTI can be viewed through three major waves:
(1) \textit{Constraint-Solving \& Logic:} Systems like IDA Pro~\cite{ferguson2008reverse,ida}, SmartDec~\cite{Fokin2011SmartDec}, Snowman~\cite{snowman}, and BinSub~\cite{smith2024binsub}, which treat type inference as a mathematical constraint problem, seeking a logically consistent ``principal type.''
(2) \textit{Sequential Neural Models:} The application of Natural Language Processing (NLP) techniques, such as Support Vector Machines (SVM) in BITY~\cite{xu2018type}, Recurrent Neural Networks (RNNs) in EKLAVYA~\cite{chua2017neural} and Convolutional Neural Networks (CNNs) in CATI~\cite{chen2020cati}, which treat assembly instructions as "sentences" to capture local usage context.
(3) \textit{Structural \& Contextual Transformers:} The latest frontier, represented by Graph Neural Networks (GNNs) like TYGR~\cite{zhu2024tygr} and Transformer-based models like SeeType~\cite{seetype}. These models capture the global "shape" of data flow and the long-range dependencies of large functions, achieving unprecedented accuracy in reconstructing fine-grained struct types.

\smallskip
\textbf{Objective of this Survey.}
Despite the significant progress in the field, there remains a lack of a unified perspective on how these diverse methodologies relate to one another. This survey provides a comprehensive analysis of the transition from heuristics to Transformers. We categorize the state-of-the-art based on their underlying representation—be it sequential, graph-based, or algebraic—and evaluate their performance across different architectures and optimization levels. By synthesizing insights from over a dozen seminal works, we identify the recurring challenges of "orphan variables" and structural reconstruction, ultimately outlining a roadmap for the next generation of "Neuro-Symbolic" type inference engines.

\section{The Anatomy of Binary Type Inference}
\begin{figure}[t]
\vspace{5pt}
    \centering
    \includegraphics[width=0.495\textwidth]{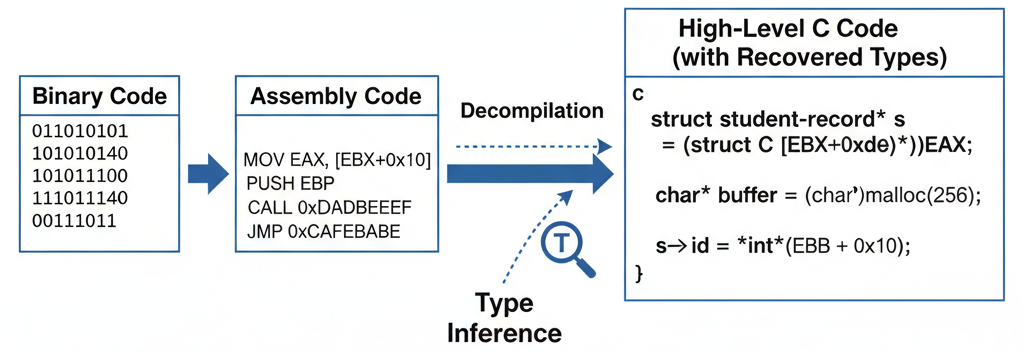}
    \setlength{\abovecaptionskip}{-2.5pt}
    \setlength{\belowcaptionskip}{2.5pt}
    \caption{Type Inference From Binary Code.}
    \label{fig:task}
\end{figure}

\begin{figure}[t]
\vspace{2.5pt}
    \centering
    \includegraphics[width=0.495\textwidth]{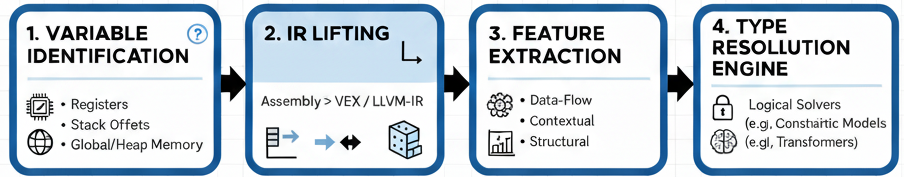}
    \setlength{\abovecaptionskip}{-2.5pt}
    \setlength{\belowcaptionskip}{-2.5pt}
    \caption{Overall Binary Type Inference Pipeline.}
    \label{fig:pipeline}
\end{figure}

To understand how high-level types are recovered from a stripped binary, one must first understand the structural chasm between source code and machine code. 
Figure~\ref{fig:task} illustrates the end-to-end workflow of this recovery process, tracing the transformation of a program from its lowest-level representation back to a human-readable high-level form.
Binary Type Inference (BTI) is essentially an inverse problem: it attempts to map a low-level, untyped execution environment back to a high-level, typed semantic space. 
This section breaks down the "anatomy" of this process into its four foundational components, as shown in Figure~\ref{fig:pipeline}: Variable Identification, Intermediate Representation (IR) Lifting, Feature Extraction, and the Type Resolution Engine.


\subsection{Variable Identification and Recovery.}
The first challenge is identifying the "entities" that require a type. In source code, variables have symbolic names and explicit scopes. In a binary, these are replaced by three types of storage locations:
\begin{itemize}
\vspace{0pt}
    \item Registers: Temporary, high-speed storage (\textit{e.g.}, \texttt{rax}, \texttt{ebp}). In optimized code, the same register may hold a pointer at the beginning of a function and an integer at the end.
    \vspace{1.5pt}
    \item Stack Offsets: Local variables stored relative to the stack pointer (\texttt{esp}/\texttt{rsp}) or base pointer (\texttt{ebp}/\texttt{rbp}).
    \vspace{1.5pt}
    \item Global/Heap Memory: Fixed addresses in the data section or dynamic addresses returned by allocators like \texttt{malloc}.
\vspace{2.5pt}
\end{itemize}

\begin{table*}[t]
\vspace{2.5pt}
\small
\centering
\setlength{\tabcolsep}{6pt}
\caption{Methodological Comparison Across BTI Research Eras}
\label{tab:methodological_comparison}
\begin{tabular}{lccc}
\toprule
\begin{tabular}[c]{@{}c@{}}\textbf{Anatomical}\\ \textbf{Component}\end{tabular}
& 
\begin{tabular}[c]{@{}c@{}}\textbf{Heuristic Era}\\ \textbf{(\textit{e.g.}, IDA/Ghidra)}\end{tabular}
& 
\begin{tabular}[c]{@{}c@{}}\textbf{Sequential Era}\\ \textbf{(\textit{e.g.}, EKLAVYA/CATI)}\end{tabular}
& 
\begin{tabular}[c]{@{}c@{}}\textbf{Structural Era}\\ \textbf{(\textit{e.g.}, SeeType/TYGR/BinSub)}\end{tabular}
\\
\midrule
Variable Recovery 
& Simple stack/register tracking 
& Metadata-based (DWARF) or basic VSA 
& Advanced program slicing and inter-procedural VSA \\
IR Representation 
& Native assembly (x86/ARM) 
& Sanitized token streams 
& Normalized VEX or P-Code IR \\
Feature Focus 
& Local mnemonic idioms 
& Local window (VUC) context 
& Global data-flow graph topology \\
Resolution Logic 
& Hardcoded If--Then rules 
& Statistical pattern matching 
& Neural attention or algebraic subtyping \\
Numeric Handling 
& Hex-only literals 
& Masked as generic \texttt{IMM} tokens 
& Numeric-aware tokenization or bitvectors \\
\bottomrule
\end{tabular}
\vspace{5pt}
\end{table*}

Tools like BITY~\cite{xu2017learning,xu2018type} and TYGR~\cite{zhu2024tygr} use Value-Set Analysis (VSA) or function-frame analysis to group these memory accesses into ``abstract variables.'' 
However, optimizations such as Frame Pointer Omission (FPO) significantly distort this process. By eliminating the stable base pointer (\texttt{ebp}/\texttt{rbp}), FPO forces the system to calculate offsets relative to a volatile stack pointer (\texttt{rsp}) that shifts with every \texttt{push} or \texttt{pop}. Consequently, a single conceptual variable may be accessed via multiple differing offsets (\textit{e.g.}, \texttt{rsp+0x10} then \texttt{rsp+0x18}), leading to fragmented variable recovery before type inference even begins.

\subsection{Intermediate Representation (IR) Lifting.}
Assembly language is architecture-specific (x86, ARM, MIPS), making it difficult to build a universal inference engine. To achieve "cross-arch" support, modern frameworks lift raw machine code into an architecture-agnostic Intermediate Representation (IR).

The VEX IR (used by angr and TYGR) or LLVM-IR simplifies complex instructions into ``micro-operations.''
For example, a single x86 push instruction is decomposed into a memory store followed by a stack pointer decrement. This normalization allows the inference engine to see the ``essence'' of the data movement—such as a 4-byte write—regardless of the CPU architecture.

\smallskip
\textbf{Feature Extraction: The ``Usage Clues''.}
Binary type inference relies on the ``Duck Typing'' principle: If it is used like a pointer, it is a pointer. The ``anatomy'' of an inference system is defined by which features it chooses to extract:
(1) Data-Flow Features: Tracking how a value from a \texttt{call} (\textit{e.g.}, \texttt{malloc}) flows into a register~\cite{khedker2017data}. If that register is later used as a base for an offset (\textit{e.g.}, \texttt{[rax + 4]}), the engine extracts a ``struct'' or ``array'' feature.
(2) Contextual Features: As highlighted in CATI~\cite{chen2020cati}, the instructions surrounding a variable provide a "Variable Usage Context" (VUC). If a variable is surrounded by floating-point arithmetic instructions, it is likely a \texttt{float} or \texttt{double}.
(3) Structural Features: TYGR~\cite{zhu2024tygr} and SeeType~\cite{seetype} extract the "shape" of data flow. By representing the function as a graph, they can see that a specific variable acts as a "sink" for multiple incoming data edges, a pattern typical of function parameters or structure fields.

\subsection{The Type Resolution Engine.}
The final component is the "brain" that maps features to labels. This engine generally falls into one of two categories:
(1) \textit{Logical Solvers}: Systems like BinSub treat types as a set of constraints. If Variable $A$ is assigned to Variable $B$, then $Type(A) \leq Type(B)$. The engine uses algebraic subtyping to find the most specific type that satisfies all constraints.
(2) \textit{Probabilistic Models}: Modern Transformer-based engines like SeeType treat type inference as a translation or classification task. The engine outputs a probability distribution (\textit{e.g.}, 85\% chance of \texttt{int*}, 10\% \texttt{void*}, 5\% \texttt{int}).

\subsection{Methodological Comparison Across Eras.}
The anatomy of BTI has shifted significantly over the decades. In the Heuristic Era, the components were rigid and reactive, resolving types only when a specific, unambiguous idiom (like a pointer dereference) was encountered. In the Sequential Era, the focus moved to linguistic context, treating assembly as a stream of tokens. Today, in the Structural Era, systems prioritize graph topology and long-range dependencies.
Table~\ref{tab:methodological_comparison} summarizes how these anatomical components are implemented across the different research generations.

\section{Paradigms of Inference: From Symbolic Logic to Neural Context}
To understand the evolution of Binary Type Inference (BTI), we must first examine the "Semantic Gap" between source code and stripped executables. We present a running example in Listings~\ref{fig:moti1}–\ref{fig:moti3} to illustrate the challenges of reconstructing high-level abstractions.

\subsection{The Semantic Gap}

Listing~\ref{fig:moti1} defines a structure \texttt{Point} and an array of integers. In the source, these are distinct types with clear boundaries. However, once compiled into the Listing~\ref{fig:asm} assembly, these abstractions vanish. For instance, the assignment \texttt{pp->y = 4} is lowered to \texttt{movl \$4, 4(\%rax)}. Without a symbol table, an automated tool cannot inherently know if \texttt{4(\%rax)} refers to a structure field, an array index, or a random stack offset.

The difficulty is further evidenced by modern decompilers. Listing~\ref{fig:moti2} (Hex-Ray of IDA Pro) fails to recover the \texttt{Point} structure, instead splitting it into two unrelated scalar variables (\texttt{v8} and \texttt{v9}). Listing~\ref{fig:moti3} (SmartDec) manages to identify the loops but labels the array pointer as a generic \texttt{void*}, losing the type-specific "scale" of the data. These failures highlight the four primary paradigms used to bridge this gap.

\subsection{Algebraic Subtyping and Symbolic Logic}

Historically, BTI was conceptualized as a constraint satisfaction problem over a formal type lattice, predicated on the assumption that while explicit type labels are erased, the implicit constraints governing data movement are preserved.

Foundational systems such as TIE~\cite{lee2011tie} and Retypd~\cite{noonan2016polymorphic} generate subtyping constraints derived from instruction behavior. For example, the instruction \texttt{mov eax, [ebx+8]} implies a constraint where \texttt{ebx} must be a pointer to an aggregate type, and the field at offset 8 must be bit-width compatible with \texttt{eax}. More recently, BinSub~\cite{smith2024binsub} reformulated this process by modeling binary types as a distributive lattice. By applying Bi-unification, BinSub reduces complex recursive constraints into simplified ``principal types,'' allowing for more efficient resolution.

While these algebraic methods offer the advantage of explainability and formal soundness, they are inherently fragile. They rely heavily on complete code coverage; in cases of ``orphan variables''—variables with sparse usage or no direct data flow to well-known library calls—the system lacks sufficient constraints to resolve a specific type, often defaulting to a generic top-type.

\begin{table*}[t]
\vspace{2.5pt}
\centering
\setlength{\tabcolsep}{10.5pt}
\caption{Comparison of Core Methodologies for Binary Type Inference}
\label{table:methodology_comparison}
\begin{tabular}{|l||l|l|l|l|}
\hline
\textbf{Metric} & \textbf{Constraint Solving} & \textbf{Sequential (RNN/CNN)} & \textbf{Graph (GNN)} & \textbf{Transformer} \\ \hline\hline
\textbf{Logic Basis} & Formal Rules & Linguistic Pattern & Structural Topology & Global Attention \\ \hline
\textbf{Accuracy} & High (if solved) & Moderate & High & State-of-the-Art \\ \hline
\textbf{Handling Sparse Data} & Poor & Good & Excellent & Excellent \\ \hline
\textbf{Function Size} & Scales well & Struggles with length & Compute intensive & Uses Slicing \\ \hline
\textbf{Primary Tools} & BinSub / Retypd & EKLAVYA / CATI & TYGR & SeeType \\ \hline
\end{tabular}
\vspace{5pt}
\end{table*}

\subsection{Sequential Modeling}

Inspired by Natural Language Processing (NLP), this approach treats binary instructions as textual sequences, deriving the meaning of a variable from the context in which it appears.

Systems such as EKLAVYA~\cite{chua2017neural} and BITY~\cite{xu2018type} pioneered the use of Word Embeddings (\textit{e.g.}, Word2Vec~\cite{church2017word2vec}) to map opcodes and operands into a high-dimensional vector space. These sequences are processed by Recurrent Neural Networks (RNNs) to identify latent type patterns. A significant advancement in this domain was the introduction of the "Variable Usage Context" (VUC) by CATI~\cite{chen2020cati}. Capitalizing on the principle of Same Type Variable Clustering—the observation that neighboring instructions tend to operate on semantically similar data—CATI utilizes Convolutional Neural Networks (CNNs) to capture local spatial locality.

While these models excel at ``intuitive'' inference for variables lacking formal constraints, they are limited by the sequential nature of their architecture. RNN-based approaches, in particular, suffer from the vanishing gradient problem, making them prone to ``forgetting'' distant instructions that may contain the only definitive clue regarding a variable's type.

\subsection{Topological Reasoning via GNNs}

Recognizing that program execution forms a complex web of dependencies rather than a linear sequence, this approach shifts focus from sequential instruction streams to rich topological graphs.

Approaches such as TYGR~\cite{zhu2024tygr} lift binary code to an Intermediate Representation (IR), such as VEX, to construct a Data-Flow Graph (DFG). In this topology, nodes represent values and edges represent operations (\textit{e.g.}, \texttt{Load}, \texttt{Store}, \texttt{Add}). Through iterative message passing, Graph Neural Networks (GNNs) update node embeddings by aggregating features from neighbors. For instance, if a node is connected to a \texttt{Dereference} operation, this structural feature propagates back to the source node, increasing the probability of it being classified as a \texttt{Pointer}. This topological approach is uniquely suited for reconstructing complex structures, as GNNs can capture the distinct ``branching'' patterns of structure member accesses more effectively than linear models.

\subsection{Transformer-Based Global Modeling}

The current state-of-the-art leverages the Transformer architecture and the Self-Attention mechanism to transcend the limitations of both local context windows and complex graph generation.

Models like SeeType~\cite{seetype} utilize the Self-Attention mechanism to model global dependencies. Unlike RNNs, which process input sequentially, Transformers compute an Attention Score between every pair of instructions in a function simultaneously. This allows the model to instantly resolve long-range dependencies, linking a variable's initialization to its usage hundreds of lines later. To maintain computational feasibility, these systems often employ Program Slicing, filtering out noise instructions to focus the model on the relevant slice of execution.

Crucially, this generation of models addresses the "Numeric Literacy" gap. Unlike predecessors that masked immediate values as generic tokens, modern Transformers incorporate Numeric-Aware Tokenization. This allows the model to interpret specific offsets (\textit{e.g.}, \texttt{0x8}, \texttt{0x10}) as explicit architectural hints for structure layout recovery rather than random noise.

\definecolor{stubbg}{HTML}{FCF3D5}
\begin{figure}[t]
\begin{lstlisting}[
		language=c,
		morekeywords={var},
		caption={The High-level Source Code of a Simple Example},
		label={fig:moti1},
		escapechar=|,
		numbers=left
	]
typedef struct {
	int x;
	int y;
} Point;

void main(){
	Point p, *pp;
	pp = &p;

	pp -> x = 3;
	pp -> y = 4;

	int a[5];
	for (int i = 0; i < 5; i++)
		a[i]=0;

	int *q = a;
	for (int i = 0; i < 5; i++)
		*(q + i) = 0;
}
\end{lstlisting}
    \vspace{15pt}
\begin{lstlisting}[
		language=c,
		morekeywords={var},
		caption={The Low-level Assembly Code of the Example},
		label={fig:asm},
		escapechar=|,
		numbers=left
	]
.LFB0:  .cfi_startproc
        endbr64
        pushq   %rbp
        .cfi_def_cfa_offset 16
        .cfi_offset 6, -16
        movq    %rsp, %rbp
        .cfi_def_cfa_register 6
        subq    $64, %rsp
        movq    %fs:40, %rax
        movq    %rax, -8(%rbp)
        xorl    %eax, %eax
        leaq    -40(%rbp), %rax
        movq    %rax, -56(%rbp)
        movq    -56(%rbp), %rax
        movl    $3, (%rax)
        movq    -56(%rbp), %rax
        movl    $4, 4(%rax)
        movl    $0, -64(%rbp)
        jmp     .L2
.L3:    movl    -64(%rbp), %eax
        cltq
        movl    $0, -32(%rbp,%rax,4)
        addl    $1, -64(%rbp)
.L2:    cmpl    $4, -64(%rbp)
        jle     .L3
        leaq    -32(%rbp), %rax
        movq    %rax, -48(%rbp)
        movl    $0, -60(%rbp)
        jmp     .L4
        ...
\end{lstlisting}
\vspace{-5pt}
\end{figure}

\definecolor{stubbg}{HTML}{FCF3D5}
\begin{figure}[t]
\begin{lstlisting}[
		language=c,
		morekeywords={var},
		caption={Code Decompiled By Hex-Ray Plugin of IDA Pro},
		label={fig:moti2},
		escapechar=|,
		numbers=left
	]
int_cdecl main(int argc, const char **argv, const char **envp)
{
	signed int j; // [sp+4Ch] [bp-30h]@4
    signed int i; // [sp+54h] [bp-28h]@1
    int v6[5]; // [sp+58h] [bp-24h]@3
    char *v7; // [sp+6Ch] [bp-10h]@1
    char v8; // [sp+70h] [bp-Ch]@1
    signed int v9; // [sp+74h] [bp-8h]@1 

    V7 = &U8;
    *(_DW0RD *)&U8 = 3;
    v9=4;
    for(i=0;i<5;++i)
        v6[i]=0;
    for(j=0;j<5;++j)
        v6[j]=0;
    return 0;
}
\end{lstlisting}
    \vspace{15pt}
\begin{lstlisting}[
		language=c,
		morekeywords={var},
		caption={Code Decompiled By SmartDec},
		label={fig:moti3},
		escapechar=|,
		numbers=left
	]
void main() {
    int32_t ebp1;
    int32_t esp2;
    int32_t v3;
    void* v4;
    int32_t v5:s
    ebp1 = esp2 -4;
    v3 = 0;
    
    while (v3<5) {
        *(int32_t*)(ebp1 +v3 * 4 + -36) = 0;
        ++v3;
    }
    v4 = (void*)(ebp1 + -36);
    v5 = 0;
    while (v5 < 5) {
        *(int32_t*)((int32_t)v4 +v5 * 4) = 0;
        ++v5;
    }
    return:
}
\end{lstlisting}
\vspace{-5pt}
\end{figure}

\subsection{Methodological Comparison Matrix}

Table~\ref{table:methodology_comparison} sthe comparative strengths and limitations of the four primary BTI paradigms. The historical trajectory reveals a distinct trade-off between formal rigor and generalization capability.
The transition across these paradigms reflects a shift from Global Logic to Local Context and finally to Global Attention. 
While symbolic logic provides a foundation of ``truth,'' neural models provide the "intuition" needed to navigate the ambiguity of optimized, stripped code.

Early constraint-solving approaches (\textit{e.g.}, BinSub~\cite{smith2024binsub}) provide a strong logical basis and high verifiable accuracy when constraints are complete; however, they remain brittle when confronted with sparse data or orphan ``variables'' where local clues are insufficient. The subsequent shift toward sequential neural models (\textit{e.g.}, EKLAVYA~\cite{chua2017neural}) introduced a probabilistic ``linguistic'' perspective, significantly improving robustness on sparse inputs but struggling to capture long-range dependencies within large functions. 
The current state-of-the-art, represented by Graph (GNN) and Transformer architectures, effectively bridges these gaps by leveraging structural topology and global attention mechanisms. 
While these modern deep learning models incur higher computational costs—often necessitating optimization strategies like program slicing—they achieve superior accuracy by analyzing the global ``shape'' of data flow rather than isolated instructions. 
Ultimately, this evolution reflects a fundamental paradigm shift: moving from the search for a provable symbolic ``truth'' to the cultivation of a statistical intuition'' capable of navigating the semantic ambiguity of optimized, stripped binaries.

\section{Cross-Cutting Challenges}
The transition from manual heuristics to high-capacity Transformer models has significantly raised the performance ceiling of binary type inference, yet the field remains embattled by the inherent ``semantic desert'' of stripped code. 
The challenges facing modern BTI are not merely artifacts of missing symbol tables; they are the result of a fundamental tension between the high-level intent of a programmer and the low-level efficiency of machine architectures. These challenges can be synthesized into four primary obstacles: the volatility of optimized code, the ``illiteracy'' of neural models regarding numerical values, the inherent ambiguity of memory layouts, and the structural integrity of the data used to train models.

\textbf{Semantics Loss Under Aggressive Optimization.}
Modern compilers prioritize execution speed and binary size over any consideration for the reverse engineer. Aggressive optimizations (\textit{e.g.}, \texttt{-O2}, \texttt{-O3}) fundamentally distort the ``anatomy'' of a program, often reordering or entirely deleting the semantic clues that human-crafted heuristics once relied upon. For instance, the widespread adoption of Frame Pointer Omission (FPO) eliminates the stable reference point provided by the base pointer (\textit{ebp}/\textit{rbp}), forcing inference engines to calculate stack offsets relative to a volatile stack pointer (\textit{esp}/\textit{rsp}) that shifts with every local operation. This creates a cascading error effect: if the variable recovery phase cannot provide a stable location, the subsequent type inference phase is essentially guessing at a moving target. Furthermore, compilers frequently reuse registers for unrelated variables across different code paths, a phenomenon known as type punning, which introduces significant label noise that confuses sequential and even some graph-based models.

\textbf{The Numeric Literacy Gap.}
A profound methodological blind spot exists in the way deep learning models handle numerical values—addresses, bitmasks, and offsets. In high-level source code, the difference between an offset of \texttt{0x8} and \texttt{0x800} is the difference between accessing a neighboring structure field and accessing an entirely different memory region~\cite{fu2024memspate}. However, to prevent a ``vocabulary explosion,'' traditional neural models have historically masked all numerical values as generic, semantically empty tokens like ``MM.'' This practice effectively blinds the model to the most critical clues for structure recovery. While a Transformer may recognize that a register is being used as a base for memory access, the loss of the specific offset prevents it from reconstructing the actual layout of the data structure. Recent innovations in numeric-aware tokenization represent a vital shift, attempting to teach models that small constants are not just numbers, but explicit hints about structure membership and alignment.

\textbf{Topological Ambiguity of Complex Types.}
While primitive types such as integers and floats are relatively easy to distinguish based on the arithmetic instructions that manipulate them, complex types like structures and unions suffer from a "topological overlap" problem. In a binary, a structure containing four 1-byte characters often looks identical to a single 32-bit integer if the compiler chooses to access the entire block via a single 4-byte move. Without the benefit of inter-procedural analysis—observing how those individual bytes are accessed in other, distant functions—the inference engine is often biased toward the simpler primitive type. This is reflected in current benchmarks: while overall type accuracy is reaching impressive heights, the accuracy for fine-grained struct member reconstruction remains a significant bottleneck. Current models can often identify a structure pointer, but they remain largely incapable of mapping the internal tree of nested members and offsets.

\textbf{Dataset Integrity and the Memorization Trap.}
Finally, the shift toward data-driven BTI has exposed a systemic flaw in how research is evaluated. The quality of any neural model is only as good as the integrity of its training data, yet the binary analysis community has struggled with a massive duplication problem. Because compilers generate identical machine code for common library functions across different versions of a software package, many standard datasets contain nearly 90\% duplicate functions. When a model is evaluated on such a dataset, its high accuracy scores are often a byproduct of rote memorization rather than a true semantic understanding of assembly logic. If the testing set contains functions that the model has effectively "seen" during training, the resulting metrics are artificially inflated. This has prompted a necessary move toward large-scale, deduplicated datasets, which force models to generalize their learning to truly novel binaries.

\section{Future Directions}
As binary type inference transitions from a niche reverse engineering task to a core component of autonomous security systems, the limitations of current deep learning models suggest a new frontier. The future of the field lies in resolving the tension between the "black-box" intuition of neural networks and the rigorous, verifiable logic of program analysis. We identify three high-impact directions that will likely define the research landscape over the next three to five years.

\textbf{Binary Foundation Models and Self-Supervision.}
Current models are largely task-specific, trained from scratch on labeled datasets to perform one specific analysis. However, the success of Large Language Models (LLMs) in natural language~\cite{wen2024automatically,ma2025Bridging,ma2026auto,su2024cfstra,wang2026g,lin2026h} suggests a move toward Binary Foundation Models. 
These are massive models pre-trained on billions of lines of unlabeled assembly code across diverse architectures and compilers using self-supervised tasks—such as Masked Language Modeling (MLM) or Jump Target Prediction.

The value of such a foundation model is its ability to learn a ``universal assembly grammar.'' Once a model understands the latent relationship between registers, memory flow, and control structures across x64, ARM, and RISC-V, it can be fine-tuned for type inference with significantly less labeled data. This approach would also address the ``cross-architecture'' challenge; a foundation model could potentially perform Zero-Shot Inference, correctly identifying types in a binary for an architecture it has never explicitly seen before by mapping the new code’s data-flow patterns into its universal semantic space.

\textit{Case study: Generative Reconstruction via GPT-5.}
The promise of this approach is best illustrated by the generative capabilities of modern LLMs like GPT-5.
When provided with a prompt, as shown in Figure~\ref{fig:prompt} containing the assembly from Listing~\ref{fig:asm} and instructions to recover the original C logic, GPT-5 produces the code shown in Listing~\ref{fig:gpt}. 
Unlike traditional decompilers that fail to group related memory accesses, the LLM demonstrates a generative leap: it ``imagines'' a plausible struct Pair definition and assigns intuitive variable names like \texttt{pair\_ptr} by reasoning about how the registers are used as base pointers. While this approach is promising due to its high readability and zero-shot ability to distinguish between direct array indexing and pointer arithmetic, it remains a ``black-box'' process prone to hallucinations. The model may synthesize code that looks idiomatic but contains subtle logic errors or violates strict machine constraints, highlighting a trade-off where semantic clarity is gained at the potential expense of formal correctness~\cite{cao-etal-2025-informal,wen2024enchanting}.

\textbf{Neuro-Symbolic Integration.}
The most prominent research gap is the lack of logical consistency in modern neural predictions. While a Transformer or GNN can provide a highly accurate "guess" based on usage patterns, these models frequently output types that are architecturally impossible or logically inconsistent with the surrounding code—such as predicting a 64-bit pointer type for a variable that only undergoes 8-bit arithmetic.

Future systems are moving toward a Neuro-Symbolic architecture~\cite{hitzler2022neuro,bhuyan2024neuro,wen2024enchanting}. In this paradigm, a neural model (like SeeType) acts as a high-capacity "generator" that proposes a probability distribution of potential types, while a symbolic logic engine (based on the algebraic subtyping found in BinSub) acts as a "verifier." By using the logic engine to prune the neural search space, we can ensure that the final output not only reflects the statistical patterns learned from millions of binaries but also adheres to the formal constraints of the C type system and machine semantics. This combination would provide both the coverage needed for ``orphan variables'' and the formal correctness required for security-critical applications like Control-Flow Integrity~\cite{ma2026integrating,jiang2025ctxfuzz}.

\begin{figure}[t]
\vspace{5pt}
\centering
\begingroup
\fcolorbox{black}{gray!10}{%
\begin{minipage}{\dimexpr\columnwidth-2\fboxsep-2\fboxrule\relax}
You are a senior reverse-engineering expert. Given the following assembly code, recover the most accurate and readable C source code possible. Infer control flow, data types, and function structure. Use meaningful names, document assumptions, and prioritize semantic correctness. Output the recovered C code followed by a short explanation of key decisions.

```

< assembly code>

'''
\end{minipage}%
}
\endgroup
\setlength{\abovecaptionskip}{5pt}
\setlength{\belowcaptionskip}{-5pt}
\caption{Prompt for translating assembly code into C Code.}
\label{fig:prompt}
\end{figure}

\definecolor{stubbg}{HTML}{FCF3D5}
\begin{figure}[t]
\begin{lstlisting}[
		language=c,
		morekeywords={var},
		caption={Generative Reconstruction via LLM (GPT-5)},
		label={fig:gpt},
		escapechar=|,
		numbers=left
	]
#include <stdint.h>

int main(void)
{
    struct Pair {
        int a;
        int b;
    } pair;

    int arr[5];
    struct Pair *pair_ptr = &pair;
    int *arr_ptr = arr;

    /* Initialize the pair via the pointer (matches stores through -56(%rbp)). */
    pair_ptr->a = 3;
    pair_ptr->b = 4;

    /* First loop: zero arr[0..4] using direct indexed addressing off rbp. */
    for (int i = 0; i <= 4; i++) {
        arr[i] = 0;
    }

    /* Second loop: zero arr[0..4] again, but via arr_ptr pointer arithmetic. */
    for (int i = 0; i <= 4; i++) {
        arr_ptr[i] = 0;
    }

    return 0;
}
\end{lstlisting}
\vspace{-5pt}
\end{figure}

\textbf{Automated Structure and Memory Layout Synthesis.}
While identifying primitive types is nearing maturity, the granular reconstruction of nested data structures remains a ``grand challenge.'' Current state-of-the-art models often collapse a complex structure into a generic ``struct'' label, losing the vital internal offsets and member relationships.

The next generation of BTI must move toward End-to-End Layout Synthesis. This involves combining static data-flow analysis with "symbolic memory snapshots." Instead of simply predicting labels, future models will likely aim to output a complete Header File (\texttt{.h}) containing reconstructed structure definitions. This requires a leap from classification to generative structural prediction, where models must account for alignment requirements, padding, and nested pointers. Integrating the topological strengths of GNNs with the long-range attentional capabilities of Transformers will be essential to capturing how a single pointer acts as the root of a recursive, complex memory tree.

\textbf{De-optimization and Semantic Normalization.}
Finally, as compiler optimizations continue to obscure binary semantics, research must explore Neural De-optimization. Rather than trying to infer types directly from ``noisy'' optimized code, a preliminary stage could use generative models to ``normalize'' or ``de-optimize'' the assembly back into a canonical form. By training models to recognize the equivalence between a complex SIMD-optimized loop and a simple C \texttt{memcpy}, we can simplify the environment in which type inference occurs. This would mitigate the impact of reordered instructions and register reuse, essentially providing the inference engine with a ``cleaner'' view of the original programmer's intent.

\section{Conclusion}
Binary type inference has evolved from a reactive process of manual heuristics to a proactive discipline of structural and neural reasoning. This survey has charted the progression from early rule-based ``duck typing'' to modern Transformer and GNN architectures that capture global data-flow context. While deep learning has significantly raised the ceiling for accuracy and structural reconstruction, the ``semantic desert'' created by aggressive compiler optimization remains a critical challenge. The next frontier lies in neuro-symbolic integration, combining the formal rigor of algebraic solvers with the statistical intuition of neural foundation models. As these paradigms converge, binary analysis will move toward a fully automated reconstruction of high-level intent, providing security analysts with a clear semantic map of stripped code.

\begin{acks}
This work was supported in part by the China Postdoctoral Science Foundation-funded project (No. 2023M723736), the Basic Research Foundation of Shenzhen City (No. JCYJ20250604184202003), the Department of Education of Guangdong Province Foundation (No. 2025KTSCX216), and Guangzhou University of Software Foundation (No. KY202412).
\end{acks}

\bibliographystyle{ACM-Reference-Format}
\bibliography{reference}

\end{document}